# On the Frequency Dependency of Radio Channel's Delay Spread: Analyses and Findings From mmMAGIC Multi-frequency Channel Sounding

Sinh L. H. Nguyen[1], Jonas Medbo[2], Michael Peter[3], Aki Karttunen[1], Katsuyuki Haneda[1],
Aliou Bamba[4], Raffaele D'Errico[4], Naveed Iqbal[5], Cheikh Diakhate[6], and Jean-Marc Conrat[6]

[1]Aalto University School of Electrical Engineering, Finland, [2]Ericsson Research, Sweden, [3]Fraunhofer HHI, Germany,
[4]CEA-LETI, France, [5]Huawei, Germany, [6]Orange, France

*Abstract*—This paper analyzes the frequency dependency of the radio propagation channel's root mean square (rms) delay spread (DS), based on the multi-frequency measurement campaigns in the mmMAGIC project. The campaigns cover indoor, outdoor, and outdoor-to-indoor (O2I) scenarios and a wide frequency range from 2 to 86 GHz. Several requirements have been identified that define the parameters which need to be aligned in order to make a reasonable comparison among the different channel sounders employed for this study. A new modelling approach enabling the evaluation of the statistical significance of the model parameters from different measurements and the establishment of a unified model is proposed. After careful analysis, the conclusion is that any frequency trend of the DS is small considering its confidence intervals. There is statistically significant difference from the 3GPP New Radio (NR) model TR 38.901, except for the O2I scenario.

*Index Terms*—5G, Frequency dependency, Delay spread, Large-scale parameters, Millimeter-wave.

## I. INTRODUCTION

In the efforts of developing an advanced and unified propagation channel model for millimeter-wave bands in the European Union Horizon 2020 mmMAGIC project [1], numerous propagation channel measurement and simulation campaigns were conducted for a variety of scenarios in the preferred suitable frequency ranges. One of the key results for developing such a model is the frequency dependency of the channel large-scale parameters (LSPs) for each type of scenario.

While frequency dependency of LSPs was investigated back in 1990s, the results have not been consistent across different studies. The previous works on frequency dependence of LSPs (mostly of the delay spread (DS)) for different indoor and outdoor environments, and for frequency bands ranging from 900 MHz up to 70 GHz, show mixed trends of increasing and decreasing DS as the frequency is higher. The frequency dependency of the DS can be seen in many measurements [2]–[10] while the others [11]–[20] support frequency independence. Any observed trend in these studies is that DS decreases with increasing frequency (except results in [18]). In general, the difference of the DS between frequency bands are small in many cases, and is only pronounced when the difference between two compared center frequencies is large, at the order of two times or more. The DS is very sensitive to the power range in the power-delay profile (PDP) applied in the calculation (i.e., DS decreases with a decrease of the power range) [13], which also impacts its frequency dependency result. It should be noted that most of these results do not fulfill requirements for comparing the DS across different measured frequencies, that are defined later in Section II. The studies that do fulfill the main requirements, i.e., [2], [13], [16], [18]–[22], show no clear frequency dependency or only slightly decreasing trend. In the recent 3GPP TR 38.901 channel model [23] for bands from 0.5 up to 100 GHz, the mean and the variance of the DS are modelled dependent on the frequency in many scenarios. Nevertheless, the relatively small number of previous works that fulfill the requirements clearly demonstrates the need for more measurements, with coordinated efforts of having similar channel sounder setups.

This paper reports the multi-frequency measurement campaigns conducted in mmMAGIC project for the frequency dependency study of the channel's LSPs, and an analysis and modelling approach enabling the evaluation of the statistical significance of the model parameters from different measurements, along with their fair synthesis to establish a unified model. Specifically, our contributions presented in this paper include:

- Providing guidelines for multi-frequency channel sounding campaigns such that their LSPs' can be compared between different frequencies;
- Providing a method to combine LSP frequency dependency models from multiple measurement campaigns into a single model by considering statistical significance of the model parameters in each campaign;
- Applying the method to DS from several multi-frequency measurements at different sites, allowing characterization of varying environmental categories of interests, i.e., outdoor, indoor and O2I; and
- Findings that frequency dependency is marginal as an overall trend after combining multiple measurement sites, while the dependency is sometimes more apparent when looking at per-site results. They are therefore significantly different from the recently reported 3GPP model, except for the outdoor-to-indoor (O2I) scenario.

The rest of the paper is organized as follows. In Section II, we first describe the list of requirements to make DSs from different channel sounding campaigns comparable. We then introduce the multi-frequency measurement campaigns conducted for this study. Section III provides the modelling methodology and discussions of the results. Finally, Section IV concludes the paper.

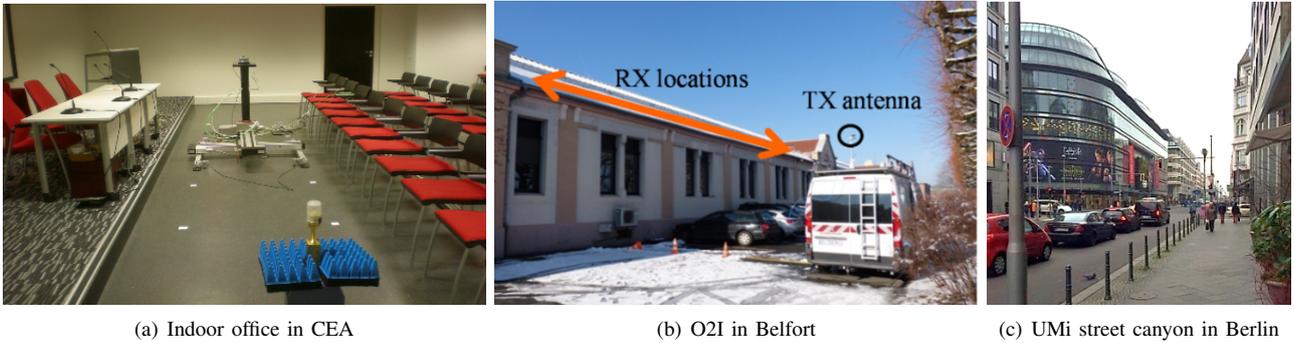

Fig. 1. Exemplary measurement scenarios in mmMAGIC.

## II. MULTI-FREQUENCY MEASUREMENT CAMPAIGNS

### A. Requirements for the channel sounding

Since these channel sounding campaigns are intended to the analysis of the frequency dependency of the various channel characteristics, the comparability across different frequencies has to be ensured. As no channel sounder is capable of covering a frequency range between 2 and 86 GHz at once, we require several channel sounder configurations operating at different radio frequencies [1], [24]. The sounders at different frequencies require specialized hardware which has a huge impact on the measurement configuration parameters, such as measurement bandwidth, dynamic range, and antenna patterns. Furthermore, it is necessary to harmonize the data post-processing methods. Several requirements have been identified (see details in Table I) that define the parameters of the channel sounding and data post-processing which need to be aligned in order to make a reasonable comparison between the different measurements taken with the different channel sounders. This list can serve as a reference in highlighting whether the conducted measurements can be combined for unified frequency dependency modeling of LSPs.

TABLE I
REQUIREMENTS FOR COMPARABILITY ACROSS DIFFERENT FREQUENCIES.

| Requirements (must be fulfilled) |
|---|
| Equal measurement bandwidth (hence the delay resolution) |
| Comparable antenna pattern, either physical or synthesized |
| Equal dynamic power range in the respective domain of analysis (e.g., delay, angle) |
| Equal spatial resolution (e.g. equal electrical array size) |
| Same environment and same antenna locations |
| **Other requirements** |
| **(small effects on the comparability or not applicable to all results)** |
| Compensation of atmospheric absorption at the 60 GHz-band |
| Sufficiently large sample size |
| Static environment (when measurements are made successively) |
| Same path estimation algorithms |
| Equal spatial averaging |

### B. Multi-frequency channel sounding campaigns

This study uses data from multi-frequency measurements conducted in 5 indoor sites, 2 O2I sites, and 5 outdoor sites in different cities in Europe by the organizations listed in Table II. All together they are equivalent to 40 single-frequency measurement campaigns. A photo of an exemplary measurement scenario for each environmental category (i.e. indoor, O2I, and outdoor) is shown in Fig. 1. Table III provides a summary of each measurement campaign including the measured frequencies, transmitter (Tx) and receiver (Rx) antenna height, link distance range, fulfilled requirements for the DS comparison between different frequencies, and the reference where the details of the conducted measurements can be found. In general, all the measurements were performed in such a way that the main requirements for comparability of the LSPs are satisfied. In several campaigns where different antenna beamwidths or bandwidths were used in the measurements, the comparability is ensured via data post-processing [1].

TABLE II
LIST OF THE ORGANIZATIONS.

| Organization | Short name |
|---|---|
| Aalto University, Finland | Aalto |
| Commissariat à l'Energie Atomique et aux Energies Alternatives, France | CEA |
| Ericsson AB, Sweden | EAB |
| Fraunhofer-Institut für Nachrichtentechnik, Heinrich-Hertz-Institut, Germany | HHI |
| Huawei Technologies Duesseldorf GmbH, Germany | HWDU |
| Orange, France | Orange |

## III. MODELLING METHODOLOGY AND RESULTS

As the measurement campaigns fulfill the requirements, the DS estimates are hence comparable between the different frequencies within each campaign. The measured root mean square (rms) DS values (in $\log_{10}$-scale, following the model in [23]) in different frequencies from the channel sounding campaigns are combined and plotted in Fig. 2 for indoor, O2I, and outdoor scenarios. The solid lines in the figures show the linear fits of each measurement campaign to the 3GPP-like model [23]:

$$\mu_{\log \text{DS}/1\text{s}} = \alpha \log(1 + f_c/1\text{GHz}) + \beta. \quad (1)$$

TABLE III
SUMMARY OF MULTI-FREQUENCY CHANNEL SOUNDING CAMPAIGNS IN MMMAGIC.

| Organization | Scenario | Tx/Rx height | Link distance range | Frequencies | Requirements for comparability | Reference |
|---|---|---|---|---|---|---|
| CEA | Indoor Office | 2.1/1.2 m | 2.3 − 7 m | 62 and 83.5 GHz | Main requirements fulfilled[†] | [1, Sec. 2.1.1] |
| CEA | Conference Room | 2.1/1.2 m | 2.7 − 8 m | 62 and 83.5 GHz | Main requirements fulfilled[†] | [1, Sec. 2.1.1] |
| EAB | Indoor Office | 1.75/1.75 m | 5 − 45 m | 2.4, 5.8, 14.8, and 58.7 GHz | All fulfilled | [1, Sec. 2.1.2] |
| Aalto | Airport | 1.6/5.0 m | 17 − 67 m | 15, 28, 60, and 86 GHz | Main requirements fulfilled[‡], sample size < 10 | [1, Sec. 2.1.3] |
| HWDU | Lecture Room | 1.68/1.68 m | 2 − 5 m | 7 and 34 GHz | All fulfilled | [1, Sec. 2.1.6] |
| Orange | O2I (Low-loss and high-loss windows) | 4/1.5 m | 10 − 25 m | 3, 10, 17 and 60 GHz | Main requirements fulfilled[†] | [1, Sec. 2.3.1] |
| EAB | O2I (Traditional building) | 1.75/1.75 m | 67 − 92 m | 2.4, 5.8, 14.8, and 58.7 GHz | All fulfilled | [1, Sec. 2.3.2] |
| HHI | Street Canyon | 1.5/5.0 m | 10 − 110 m | 10.25, 28.5, 41.5, and 82.5 GHz | All fulfilled | [1, Sec. 2.2.6] |
| EAB | Street Canyon | 1.75/1.75 m | 67 − 92 m | 2.4, 14.8, and 58.7 GHz | All fulfilled | [1, Sec. 2.2.3] |
| Aalto | Street Canyon | 1.9/1.9 m | 10 − 60.5 m | 15, 28, 60, and 86 GHz | Main requirements fulfilled, sample size < 10 | [1, Sec. 2.2.1] |
| Aalto | Open Square | 1.6/5.0 m | 10 − 60 m | 28 and 83 GHz | Main requirements fulfilled, | [1, Sec. 2.2.4] |
| Orange | Street Canyon | 4/1.5 m | 16 − 200 m | 3, 17, and 60 GHz | Main requirements fulfilled[†] | [1, Sec. 2.2.10] |
| Orange | Open Square | 4/1.5 m | 16 − 200 m | 3, 17, and 60 GHz | Main requirements fulfilled[†] | [1, Sec. 2.2.10] |

[†] Different antennas were used in the measurements, but omnidirectional patterns were synthesized in post-processing
[‡] Different bandwidths were used in the measurements, but the same bandwidth of 2 GHz was used in post-processing.

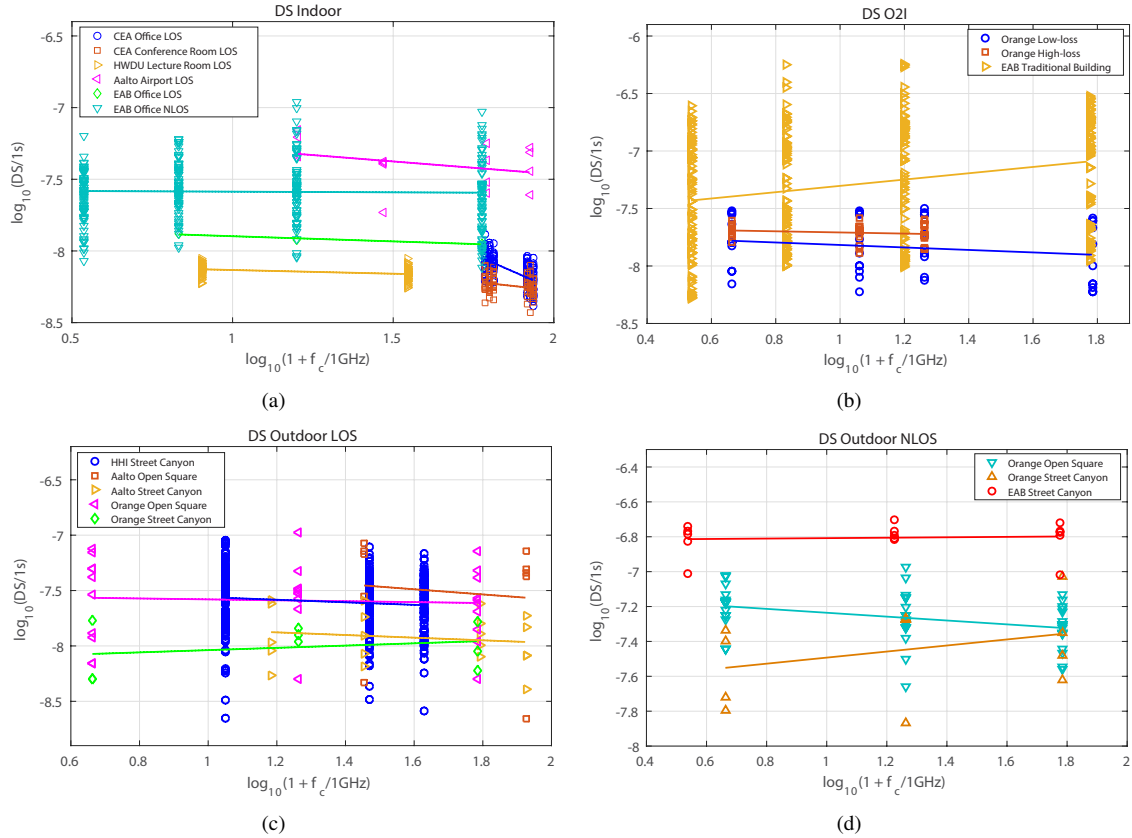

Fig. 2. DS versus frequency (logarithmic units) in (a) indoor, (b) O2I (c) outdoor LOS and (d) outdoor NLOS multi-frequency measurement campaigns. The solid lines show the linear fits.

TABLE IV
FREQUENCY-DEPENDENT LINEAR REGRESSION MODEL FOR THE DS IN EACH CAMPAIGN[1]

| Scenario | | | Model: $\mu_{\log \text{DS}/1s} = \alpha \log(1 + f_c/1\text{GHz}) + \beta$ | | |
|---|---|---|---|---|---|
| | | | $\alpha$ (95% confident bounds) | $\beta$ | $p$-value |
| Indoor | LOS | **CEA Office** | **−0.98 (−1.17, −0.78)** | **−6.31** | **0.000** |
| | | **CEA Conference Room** | **−0.24 (−0.46, −0.03)** | **−7.79** | **0.025** |
| | | **HWDU Lecture Room** | **−0.05 (−0.08, −0.03)** | **−8.08** | **0.000** |
| | | Aalto Airport | −0.18 (−0.47, 0.11) | −7.11 | 0.207 |
| | | EAB Office | −0.07 (−0.27, 0.12) | −7.83 | 0.134 |
| | NLOS | EAB Office | −0.01 (−0.06, 0.04) | −7.58 | 0.735 |
| O2I | | Orange Low-loss | −0.11 (−0.29, 0.07) | −7.71 | 0.231 |
| | | Orange High-loss | −0.05 (−0.14, 0.04) | −7.66 | 0.286 |
| | | **EAB Traditional building** | **0.27 (0.16, 0.39)** | **−7.58** | **0.000** |
| Outdoor | LOS | **HHI Street Canyon** | **−0.12 (−0.226 − 0.005)** | **−7.45** | **0.041** |
| | | Aalto Open Square | −0.23 (−2.00, 1.53) | −7.12 | 0.768 |
| | | Aalto Street Canyon | −0.12 (−0.50, 0.26) | −7.73 | 0.521 |
| | | Orange Open Square | −0.05 (−0.32, 0.22) | −7.53 | 0.773 |
| | | Orange Street Canyon | 0.10 (−0.28, 0.49) | −8.14 | 0.549 |
| | NLOS | **Orange Open Square** | **−0.11 (−0.20, −0.02)** | **−7.13** | **0.014** |
| | | Orange Street Canyon | 0.13 (−0.20, 0.47) | −7.57 | 0.294 |
| | | EAB Street Canyon | 0.01 (−0.09, 0.12) | −6.82 | 0.803 |

In each campaign, the same dynamic range (20 or 25 dB, depending on the campaign) was used to calculate the DS values for different measured frequencies to have a reasonable comparison. Table IV presents the model parameters for each campaign from the linear regression of the DS over radio frequencies and the accompanied 95% confidence bounds of the slopes $\alpha$, the $p$-values for the hypothesis that there is a frequency trend ($\alpha \neq 0$). The campaigns highlighted with the boldfaced text are the ones where the frequency trend is statistically significant with 95% confidence level, i.e., the $p$-value is small enough (smaller than 0.05 in this study) to reject the null hypothesis that "The DS has no increasing or decreasing linear relationship with the frequency".

As can be seen from Table IV, the channel sounding campaigns show differences in the trend of frequency dependency and with varying levels of statistical significance. There is no obvious and clear trend amongst the different campaigns. It should be noted that though the total number of measured DS samples is large, the number of investigated realizations of each environmental category, i.e., outdoor, indoor, and O2I is still limited. The corresponding observed frequency trends may therefore be due to the specific characteristics of each scenario and environment. To assess more general trends, the DS samples from similar scenarios (indoor office, street canyon, O2I) have been merged and analyzed together for LOS and NLOS conditions. The corresponding model parameters are obtained by weighted averaging of those from each measurement campaign. For the slope $\alpha$, it is given by:

$$\overline{\alpha} = \frac{\sum_{i=1}^{N} w_i \alpha_i}{\sum_{i=1}^{N} w_i}, \quad (2)$$

where $w_i = 1/\sigma_i^2$, $\sigma_i^2$ is the variance of the estimated $\alpha_i$ in the linear regression for the $i$-th campaign. The 95% confidence bounds of the combined slope are calculated as

$$[\overline{\alpha} - 1.96\overline{\sigma}, \overline{\alpha} + 1.96\overline{\sigma}], \quad (3)$$

where the variance of the combined model is:

$$\overline{\sigma}^2 = \left(\sum_{i=1}^{N} \frac{1}{\sigma_i^2}\right)^{-1}. \quad (4)$$

The same calculation is used for weighted intercept point $\overline{\beta}$.

TABLE V
FREQUENCY-DEPENDENT LINEAR REGRESSION MODEL FOR THE DS IN EACH MERGED SCENARIO[1]

| Scenario | | Weighted $\overline{\alpha}$ (95% confidence bounds) | Weighted $\overline{\beta}$ |
|---|---|---|---|
| Indoor Office* | LOS | −0.05 (−0.07, −0.037) | −8.03 |
| | NLOS | −0.01 (−0.06, 0.04) | −7.58 |
| Indoor Airport** | LOS | −0.18 (−0.47, 0.11) | −7.11 |
| O2I | | 0.05 (−0.02, 0.11) | −7.64 |
| Street Canyon | LOS | −0.11 (−0.19, −0.02) | −7.50 |
| | NLOS | 0. (−0.07, 0.11) | −6.89 |
| Open Square** | LOS | −0.06 (−0.34, 0.23) | −7.52 |
| | **NLOS** | **−0.1 (−0.20, −0.02)** | **−7.13** |

* The CEA measurements are not included in the combined fit due to the limited relative frequency range.
** The scenario is not covered by the current 3GPP model.

The model parameters in each merged scenario, i.e. taking into account all the measurement campaigns belonging to the environmental category of interests, are presented in Table V. The results show that the overall trend, with 95% statistical confidence level, that the DS decreases slightly for Indoor LOS, Street canyon LOS, and Open Square NLOS scenarios, as the frequency increases. For the other scenarios, any frequency trend is upper bounded by the confidence range. Corresponding maximum slope values are −0.06, −0.47, 0.11, 0.11, −0.34 for Indoor Office NLOS, Indoor Airport LOS, O2I, Street Canyon NLOS and Open Square LOS scenarios, respectively. It may thus be concluded that only small frequency dependencies are statistically likely, suggested by the confidence intervals, except for the Indoor Airport LOS and Open Square LOS scenarios. The main conclusion is therefore that the frequency dependency of the DS values from mmMAGIC measurements is small in general.

[1]The boldfaced rows indicate that the frequency trend found in the corresponding campaign is statistically significant with 95% confidence level.

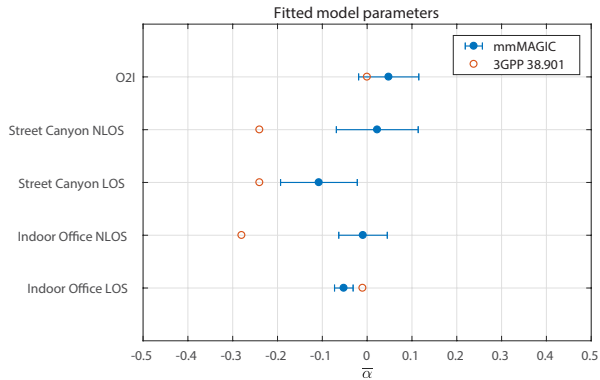

Fig. 3. Fitted $\overline{\alpha}$ and corresponding confidence intervals (95%) for the different measurement campaigns.

In Fig. 3 the slopes of the frequency dependency model for DS are compared with the 3GPP model [23]. Generally, 3GPP model results in a slope with a more negative value as compared to ours. This is clear difference between the two models, perhaps due to the different data combining methods and different source of measurements in the 3GPP model.

IV. CONCLUSION

We have provided guidelines and requirements for the multi-frequency channel measurements to analyze the frequency dependency of the LSPs, and an analysis and modelling method to combine results from different measurement sites. By applying the method to study the frequency dependency of DS values obtained from mmMAGIC multi-frequency measurement campaigns in different environments, we have found that there is no clear linear frequency dependency of the DS considering the confidence intervals. In general, the trend is small and largely dependent on the specific scenario.

ACKNOWLEDGEMENT

The authors would like to thank colleagues in the mmMAGIC WP2 for their contributions to this paper. The research leading to these results received funding from the European Commission H2020 programme under grant agreement No 671650 (5GPPP mmMAGIC project).